\documentclass[a4paper, 12pt]{spieman}  
\usepackage{tocloft}
\usepackage{soul} 
\usepackage{comment} 
\usepackage{amsmath,amsfonts,amssymb}
\usepackage{graphicx}
\usepackage{setspace}
\usepackage{placeins}
\usepackage{upgreek}

\usepackage{draftwatermark}
\SetWatermarkText{Preprint}
\SetWatermarkScale{1.5} 
\SetWatermarkColor[gray]{0.85} 
\SetWatermarkAngle{45}

\title{Blob Detection for Photonic Metasurface Designing: Angular and Spectral Control of Scattered Light}

\author[a]{Anja Tiede}
\author[b,c]{Nick Feldman}
\author[b,d]{Alexander Lambertz}
\author[b,d]{Femius Koenderink}
\author[a,e]{Anna Fontcuberta i Morral}
\author[b,f, *]{Esther Alarcon-Lladó}
\affil[a]{Laboratory of Semiconductor Materials, Institute of Materials, EPFL, 1015 Lausanne, Switzerland}
\affil[b]{Center for Nanophotonics, AMOLF, Science Park 102, 1098 XG Amsterdam, The Netherlands}
\affil[c]{ARCNL, Science Park 106, 1098 XG Amsterdam, The Netherlands}
\affil[d]{WZI, University of Amsterdam, Science Park 904, 1098 XH Amsterdam, The Netherlands}
\affil[e]{Institute of Physics, EPFL, 1015 Lausanne, Switzerland}
\affil[f]{HIMS, University of Amsterdam, Science Park 904, 1098 XH Amsterdam, The Netherlands}

\cftpagenumbersoff{figure}
\cftpagenumbersoff{table}

\begin{document} 
\maketitle

\begin{abstract}
Metasurfaces with precise spectral and angular control of light scattering are of growing interest for photonic applications requiring advanced photon management.
Correlated disorder emerged as a promising route for angular control of light scattering, but most design approaches are computationally expensive or do not allow spectral tunability.
Here, we introduce a reverse-engineering design approach for correlated-disordered metasurfaces based on tailoring the Fourier space and blob detection to create point distributions that are later "decorated" with individually designed nano-resonators or meta-atoms to tune the optical response for the desired functionality.
We validate the control of the angular scattering by fabricating ensembles of Au nanopillars following our design and characterizing their angular scattering with Fourier microscopy.
Using finite-difference time-domain simulations, we demonstrate how the choice of decorative unit tunes the spectral response, showcasing individual and independent control over light scattering in angular and spectral terms. Lastly, we expand the limits of our versatile approach by combining multiple metasurfaces in one, effectively adding their individual scattering characteristics.
As such, we can address spectral and angular ranges independently, yielding a high degree of control of the scattering response.
\end{abstract}


{\noindent \footnotesize\textbf{*}Esther Alarcon-Lladó,  \linkable{e.alarconllado@amolf.nl}}

\begin{spacing}{2}   

\section{Introduction}
\label{sect:intro}  

Metasurfaces have garnered significant attention because they can manipulate light scattering across a broad angular and spectral range.\cite{neder_combined_2019, haghtalab_ultrahigh_2020, koirala_highly_2018, yu_flat_2014}  
Moreover, metasurfaces composed of metaatoms arranged in a periodic structure enable arbitrary wavefront shaping as the individual metaatom can be chosen to adapt the ensemble's response for different functionalities \cite{balthasar_mueller_metasurface_2017, arbabi_dielectric_2015, whiting_meta-atom_2020, liu_multipole_2020}. 
Particularly, metasurfaces with a certain degree of disorder offer enhanced broadband light trapping capabilities because of their optimized k-space, making them highly effective for thin film solar cells \cite{tavakoli_over_2022, van_lare_optimized_2015, yu_design_2017, lee_concurrent_2017, buencuerpo_engineering_2021, martins_deterministic_2013}. 
Typical design procedures for these light-trapping structures include inverse design routines and/or genetic algorithms to solve an optimization problem. 
While recent reports further expand to machine learning-based and neural network-assisted design approaches, these remain computationally intensive \cite{ji_recent_2023, molesky_inverse_2018}.

A more elegant solution is to design metasurfaces directly in k-space using a 1D representation function, the spectral density function (SDF), resulting in interconnected channel-type structures \cite{yu_design_2017, lee_concurrent_2017}.
The SDF is the radially averaged power spectral density. It represents the structure in reciprocal space and characterizes its light-matter interaction because of the relation between Fourier space and far-field scattering.
This concept has been widely utilized to obtain light trapping patterns for solar cells using either iteration-based routines (optimization \cite{van_lare_optimized_2015, martins_deterministic_2013, buencuerpo_engineering_2021}, simulated annealing \cite{chen_designing_2018}), naturally occurring spinodal decomposition processes \cite{buencuerpo_efficient_2022, lee_concurrent_2017} or design-based techniques \cite{yu_design_2017, tavakoli_over_2022}.
Unfortunately, each approach yields interconnected channel-type structures that are limited in design variability because they do not have isolated decoration units with individually tunable responses, e.g., by varying their size or shape \cite{van_de_groep_designing_2013, butakov_designing_2016, lukyanchuk_optimum_2015, rahimzadegan_comprehensive_2022}.
Exploiting this can enable multi-resonant responses, desirable for, e.g., enhancing light coupling in thin film solar cells \cite{spinelli_light_2014, lee_multiresonant_2023}.
Various k-space engineering approaches have been proposed to design isolated structures with a determined k-space, for example, using random sphere packing \cite{yu_design_2017} or molecular-dynamics simulated annealing procedures \cite{froufe-perez_role_2016}.
However, these methods offer limited flexibility in designing the individual decoration unit and/or rely on computationally expensive iterative routines.

This paper introduces and experimentally demonstrates an efficient design approach to generating a metasurface consisting of isolated metaatoms with a targeted k-space. 
We propose to combine blob detection, an image analysis concept, with k-space engineering to identify locations with targeted spatial correlation.  
A second step is to re-decorate the points with meta-atoms.
This approach allows flexible tunability through designing the individual scattering units: 
Tailoring resonances in the decoration units can engineer the spectral response, while the correlation in the spatial distribution of the scatterers determines the angular scattering of light.
We show that it is also possible to design more complex angular and spectral scattering properties: By interlacing different basic elements into a single metasurface, one can obtain a scattering response that is the sum of the responses of the two basic structures. 
Taking these qualities together, our proposed approach provides a highly flexible toolbox to design photonic metasurfaces.

\section{Materials and Methods}
\label{sect:matmet}

\subsection{Input and output patterns representation}

This work uses the spectral density function (SDF) to characterize the real-space patterns, a concept readily employed in microstructure analysis and photonic design \cite{yu_design_2017, yu_characterization_2017, iyer_designing_2020, shi_three-dimensional_2025, lee_concurrent_2017}. 
The 1D SDF is the normalized azimuthally-averaged magnitude square of the Fourier spectrum of a real space structure $Z(\mathbf{r}) $:

\begin{equation}
\begin{split}
F\{Z(\mathbf{r})\} (\mathbf{k})= A_\mathbf{k} \cdot e^{i\phi_\mathbf{k}} \\
f_\mathrm{SDF}(\mathbf{k}) = (A_\mathbf{k})^2 / C,
\end{split}
\label{eq:SDF}
\end{equation}
with $A_k$ and $\phi_k$ the magnitude and phase values at each point $k$ in reciprocal space and $C$ a normalization constant.
We only consider in-plane isotropic diffraction; thus, $f_\mathrm{SDF}(\mathbf{k})$ can be reduced to a 1D scalar representation $f_\mathrm{SDF}(k)$ of the 2D Fourier space.
The SDF contains information on structural correlations in the real space pattern. This is similar to, e.g., X-ray diffraction, where the far field diffraction pattern reveals the structure factor, i.e., the Fourier transform of the real space pattern at hand.
Moreover, the SDF characterizes the angle-dependent light scattering of the metasurface, and therefore, we use it as input for the design routine. 
From this, we generate a point pattern and assign a decoration unit to each point (see next section), creating a real space pattern with isolated unit cells that contains most of the structural correlations of the input design. 
In the first Born approximation (i.e., the limit of single scattering that applies to low index-contrast structures), this corresponds to evaluating the product of the structure factor (i.e., the Fourier transform of the lattice geometry) and the form factor (i.e., the single unit cell scattering function approximated as the Fourier transform of the geometry).

\subsection{Point pattern generation with blob detection on correlated Gaussian random fields}

From the input SDF, we generate binary patterns using the Gaussian random field (GRF), also known as the spinodal approach \cite{teubner_level_1991,yu_design_2017,ma_random_2017, lee_concurrent_2017}.
To construct the (complex) Fourier space from the 1D SDF, we assign each point an amplitude $A_k$ proportional to the corresponding point on the SDF, and a random phase $\phi_k$ uniformly distributed in the range $[-\pi,\pi)$.
The real-space Gaussian random field $\Phi(\mathbf{r})$ is obtained via an inverse Fourier transform; pixel values in $\Phi(\mathbf{r})$ follow a Gaussian distribution. 
Many works use the GRF-generated real-space function ($\Phi(\mathbf{r})$) to obtain a binary pattern by thresholding $\Phi(\mathbf{r})$ with a defined cut-off value ($\Phi_0$) so that all positions $r$ with $\Phi(\mathbf{r}) < \Phi_0$ are assigned to one material and those where $\Phi(\mathbf{r}) > \Phi_0$ are assigned to the second material. 
$\Phi_0$ is a value chosen to yield a desired filling fraction. Thresholding yields a spinodal binary pattern with interconnected features where only minor spatial correlations are lost compared to $\Phi(\mathbf{r})$.  

Instead of thresholding, in this work, we transform the real-space function $\Phi(\mathbf{r})$ to generate a point pattern with isolated structures using blob detection. 
In image analysis, a "blob" is defined as a region in an image that visually differs from its local surroundings \cite{koenderink_structure_1984, lindeberg_scale-space_1994}.
Blob detection is an established method in image processing that has found its way into disciplines from medicine (e.g., to identify caries \cite{majanga_automatic_2021} or cancer \cite{di_ruberto_blob_2020}) to physics (e.g., to identify atomic columns in transmission electron microscopy images \cite{manzorro_exploring_2022}). 
Specifically, we use the Laplacian of Gaussians variant from the scikit-image Python library because of its higher read-out accuracy \cite{van_der_walt_scikit-image_2014} to identify a set of coordinates with similar spatial correlations as in $\Phi(\mathbf{r})$. Full implementation details are provided in the Supplementary Information file.

\subsection{Au nanopillar fabrication}
We will show that while the spatial correlation of the points controls the angular scattering, spectral features are determined by the scatterers that we place at the identified points in the point pattern.
In this work, we report on Au nanopillars as scatterers which we use because of their strong resonant scattering that can be tuned via the Au nanopillar dimensions.
Au nanopillar metasurfaces were fabricated on a silicon substrate using e-beam lithography (Raith system) and MMA/PMMA positive resist, followed by evaporation and lift-off.
The size of each pattern was 2~$\times$~2~mm$^2$, realized by repeating a 50~$\times$~50~$\upmu$m$^2$ unit cell. 
$\sim$~300~nm thick Au was evaporated on top of a 10~nm Ti wetting layer, followed by lift-off in acetone. 
The pillar height was confirmed using atomic force microscopy. 
The actual diameters of the pillars differ from the input design and were confirmed by scanning electron microscopy. 
No differences in pillar geometries were found between the center and edge of the pattern.

\subsection{Fourier microscopy}

We use Fourier microscopy in reflection mode to identify the angular-dependent scattering response of the patterns. 
Fourier microscopy allows us to measure the angular dependence of light scattering by imaging on the back focal plane \cite{cueff_fourier_2024}.
The principle is described in more detail here \cite{sersic_fourier_2011, osorio_k-space_2015, mohtashami_angle-resolved_2015}.
We illuminate our samples with a supercontinuum laser (NKT Whitelase Micro) coupled to a single-mode optical fiber and collimated to a Gaussian beam. 
The illumination wavelength is selected with a 10 nm band-pass filter centered around 640 nm (Thorlabs) and the total power of the nominally unpolarized light is 1.145~$\upmu$W.
The scattered light is captured by the objective (NA=0.95) and then imaged onto a C-MOS camera (Basler, acA1920-40um) using a relay lens setup.
Each image is taken with 20~ms integration time.
A beam block placed in the conjugate Fourier plane of the objective reduces the influence of the directly reflected light from the sample in the image.


\subsection{Electromagnetic simulations}
To complement our study on plasmonic realizations, we also explore dielectric nanoscatterer point patterns.
To that end, we implement a calculation workflow in which far-field light scattering and forward scattering cross-sections were calculated using a commercial finite-difference time-domain (FDTD) solver \cite{lumerical_fdtd}. 
The patterns are considered a dielectric with n~=~3.5 and height h~=~200~nm on a lossless, dispersion-less substrate of the same dielectric.
For the far-field light scattering calculations, the simulation size was 10~$\times$~10~$\times$~0.38~$\upmu$m$^3$ with a minimum mesh step size of 6~nm using automatic non-uniform meshing conditions and a refined mesh of 2~nm step size around the pattern.
Periodic boundary conditions were implemented in $x$ and $y$ directions and perfectly matched layers in the $z$ direction.  
To quantify the power scattered per unit angle, incident power, and photon energy, the far field projection is calculated from a transmission monitor 20~nm below the pattern.
Due to the high memory requirements of the simulations, they are performed on a supercomputer cluster.

For the scattering cross-section calculations, the simulation size is chosen as 1.2~$\upmu$m in all dimensions with a minimum mesh step size of 2.5~nm and using conformal meshing.
The total-field-scattered-field source is injected along the $z$ axis, spana 500~nm in all directions, and covers the wavelength range from 300~nm to 1100~nm.
Field monitors are placed only within the substrate to register the fields scattered into the substrate.
We calculate the forward scattering cross-section $Q_\mathrm{scat}$ by normalizing the transmitted power through the monitor to the source intensity and the geometrical cross-section of the scattering object.

\section{Results and Discussion}
\label{sect:res}

\subsection{Non-iterative approach for generating correlated-disordered point patterns using blob detection}

\FloatBarrier

\FloatBarrier
\begin{figure}[h!]
    \centering
    \includegraphics{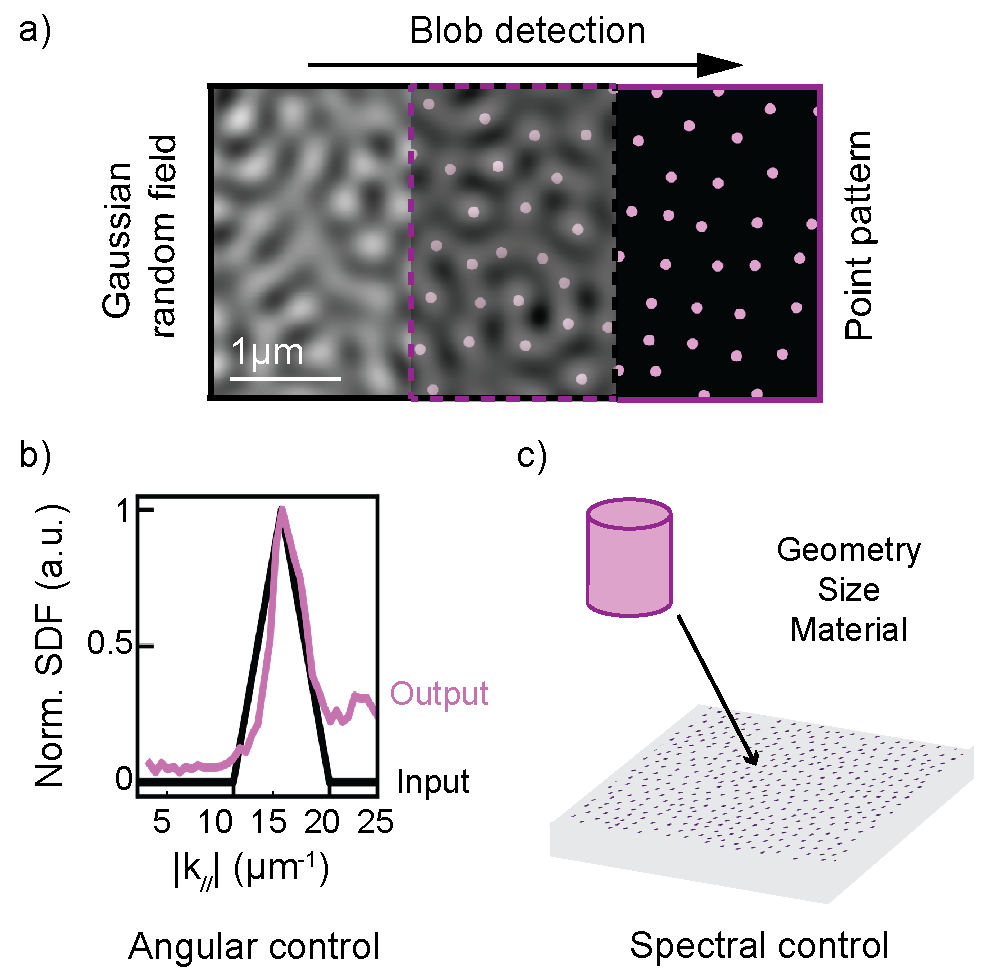}
    \caption{Metasurface generation using Gaussian random field modeling, blob detection, and nano-resonator decoration.
    The point pattern provides the angular control of light scattering as highlighted in (a,b).
    (a) Correlated Gaussian random field image (left) and point pattern identified by the blob detection algorithm (right). The middle area is an overlay of the two images.  
    (b) 1D representation function of the Fourier space through the spectral density function (SDF)  of the input (black line) and extracted from the generated pattern (purple line).
    (c) Sketch of a metasurface obtained by decorating the point pattern with nano-resonators. The nano-resonator's geometry, size, and material control the spectral response.
    }
    \label{fig:fig0}
\end{figure}

We introduce a non-iterative inverse-design approach to generate two-dimensional point patterns, which serve as a foundation for metasurfaces with well-defined angular and spectral scattering responses (see Fig. \ref{fig:fig0}a-c). The angular scattering is governed by the correlated-disordered lattice in real space, while the spectral response is determined by the geometry of nanoscatterers used to decorate the identified positions in the point pattern.

The correlated-disordered point pattern is generated from the momentum space through a combination of Gaussian random field (GRF) modelling \cite{yu_design_2017, lee_concurrent_2017} and blob detection (see Methods and Supplementary). 
We use the spectral density function (SDF) as a 1D representation of the desired power distribution in the momentum space (black line in Fig. \ref{fig:fig0}b) to generate a GRF image in the real space (Fig. \ref{fig:fig0}a-left). 
A Laplacian of Gaussians blob detection algorithm is applied to identify high-correlation regions within the GRF, resulting in a set of spatial coordinates (Fig. \ref{fig:fig0}a-right). 
These points correspond to local maxima in intensity and align with regions of enhanced spatial correlation in the original GRF.
In other words, the point pattern retains most structural correlation prescribed by the input SDF, which is also evident from the close match between the SDF of the final pattern (purple line) and the input (black line) in Fig. \ref{fig:fig0}b.
This property is true for a wide range of input SDF, making our approach as versatile as GRF modelling. A more detailed comparison is provided in the Supplementary Information.
Notably, our approach requires no iterative loops and scales robustly with system size, defined by the unit cell size and resolution, so it is only limited by the available computational memory.

Once the 2D point pattern is generated, we decorate each point in real space with a scattering unit. 
This method is an established approach for the design of metasurfaces with periodic lattices.
Each decoration unit can be modified in geometry, dimensions, or material to design the individual response, thereby tuning the ensemble's response \cite{rahimzadegan_comprehensive_2022, lukyanchuk_optimum_2015,butakov_designing_2016}.
More recently, it has been extended to hyperuniform correlated-disordered systems \cite{froufe-perez_role_2016, tavakoli_over_2022}.
In this work, we modify the decoration unit's spectral scattering characteristics to demonstrate that in our approach, the point pattern generation and the resonator design present two separate parameters that allow independent angular and spectral control tailoring.

\FloatBarrier

\subsection{Demonstration of angular scattering control}
\begin{figure}
\begin{center}
\begin{tabular}{c}
\includegraphics[width=\textwidth]{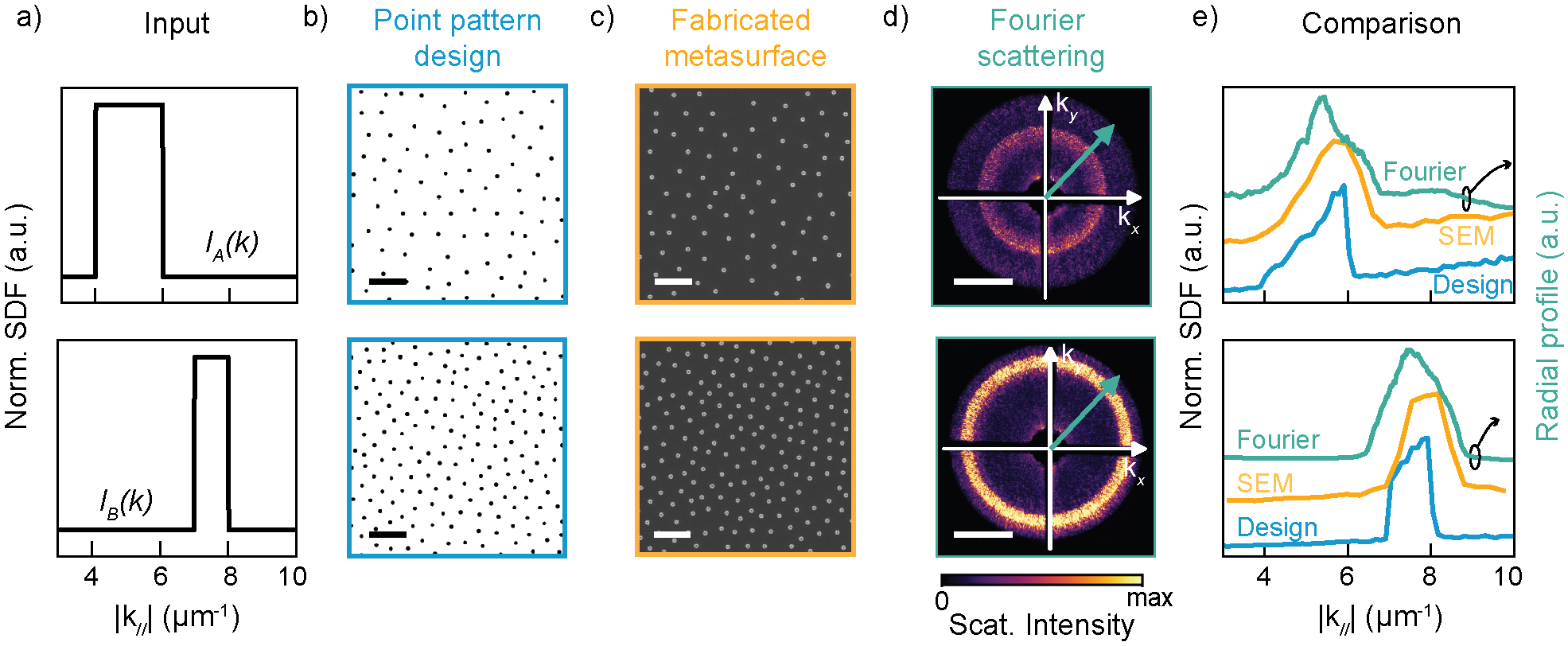}
\end{tabular}
\end{center}
\caption 
{\label{fig:fig6}
Correlated-disordered metasurface demonstration. 
(a) Two rectangular SDF profiles, $I_A(k)$ and $I_B(k)$, are used as input functions. The SDFs have a plateau for $k_A \in [4,6]~\upmu$m$^{-1}$ and $k_B \in [7,8]~\upmu$m$^{-1}$.
(b) Real space point patterns created from the input functions in a).
(c) Scanning electron microscopy (SEM) images of the fabricated metasurfaces with Au nanopillars ($r\sim$130~nm, $h\sim$300~nm) on Si. Scale bars are 2~$\upmu$m for (b,c).
(d) Fourier scattering microscopy images at $\lambda$~=~640~nm from each metasurface. The colorbar indicates the scattering intensity, and both measurements are normalized to the common maximum. Scale bar corresponds to 5~$\upmu$m$^{-1}$.
(e) SDFs of the real space design (blue lines), SEM images (orange lines), and Fourier reflectance images (green lines). The graphs are normalized to the maximum value in the depicted range and shifted on the vertical axis for clarity.
} 
\end{figure} 

Having introduced the method, we now experimentally examine the angular scattering of an ensemble of nanopillars placed according to our design approach. 
To this end, we fabricate two metasurfaces placing Au nanopillars at the coordinates given by two rectangular input functions: $I_A(k)$ and $I_B(k)$, where $I_A(k)$ peaks between $k\in [4,6]~\upmu$m$^{-1}$ and $I_B(k)$ between  $k\in [7,8]~\upmu$m$^{-1}$ (see Fig. \ref{fig:fig6}a). Figure \ref{fig:fig6}b shows the resulting point patterns. 
As expected from the input, the point density is higher for input functions with larger spatial frequencies. 
Fig. \ref{fig:fig6}c shows representative scanning electron microscopy (SEM) images of the as-fabricated metasurfaces. 
In both patterns, the fabricated nanopillars have a diameter of $\sim$~260~nm and a height of $\sim$~300~nm. 

We use Fourier microscopy to characterize the metasurfaces at $\lambda$~=~640~nm (see Fig. \ref{fig:fig6}d) by illuminating the sample with unpolarized light through normal incidence wide field illumination.
Fourier microscopy gives insight into the angular scattering response of the structures in a single measurement by imaging the scattered signal in the back focal plane. 
The black cross through the center is caused by a beam block that blocks most of the zero-order reflection.
Details on the measurement setup can be found in Methods.
In both angular scattering patterns, isotropic scattering can be observed as a ring around the center. 
As expected, the ring radius is smaller in the metasurface generated from $I_A(k)$ than that generated by $I_B(k)$. 
It is interesting to note that the metasurface with the higher point density shows a larger scattering strength.
Given that the pillars' dimensions are identical in both patterns, we expect that the scattering intensity is modified solely by the density of pillars in each design. 
To better compare the scattering profile with the input SDF, Fig. \ref{fig:fig6}e shows the radially-averaged profiles of the Fourier reflectance image (green) and SDF of the generated point pattern (blue), as well as the radially averaged Fourier transform of the SEM image (orange).
Each curve is normalized to its maximum, and the curves are vertically shifted for clarity.  
The three curves agree very well with each other and the input functions, supporting the claim that the correlated real-space locations govern the angular light scattering.

\FloatBarrier

\subsection{Defining the spectral response by choice of decoration unit}

\begin{figure}
\begin{center}
\begin{tabular}{c}
\includegraphics[width=\textwidth]{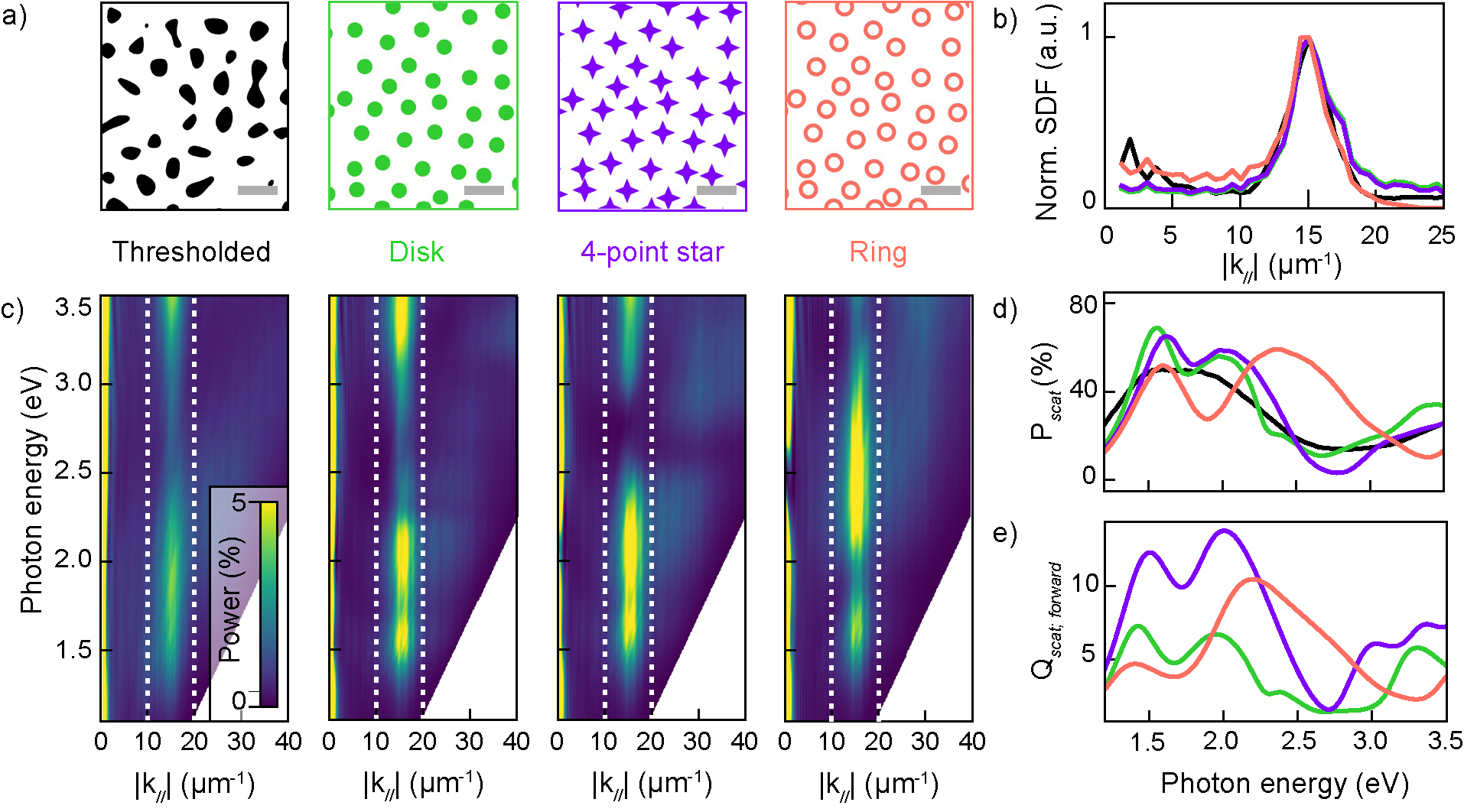}
\end{tabular}
\end{center}
\caption 
{ \label{fig:fig3} 
Comparison of FDTD results for scattering into the far-field through a pattern decorated with differently shaped scattering units and forward scattering of each single scattering unit. 
The same triangular symmetric input function creates four patterns in (a): a thresholded version and one set of coordinates for the point patterns. The points are decorated with different shapes, namely rings (red), circles (green), and stars (purple). The dimensions of the single resonators and the binary threshold were chosen to maintain a filling fraction of around 16\% in the full pattern. The scale bar is 0.5~$\upmu$m.
(b) SDF of the patterns in (a). 
(c) Angular dependence of the scattering for different patterns calculated from the far-field transformation of the power transmitted through the indicated pattern (dielectric n~=~3.5, h~=~200nm) on a lossless, dispersion-less substrate of the same material. The color represents the fraction of power scattered into a specific in-plane momentum range.
White dashed lines indicate $k_1~=~10~\upmu$m$^{-1}$ and $k_2~=~20~\upmu$m$^{-1}$, the region where the scattering of the patterns is designed for.
(d) Integrated power scattered into the region of $[k_1, k_2]$.
(e) Forward scattering cross-sections of each resonator (h~=~200~nm, dielectric of n~=~3.5 placed on a substrate of the same refractive index).
} 
\end{figure} 

While the angular scattering response is governed by the spatial arrangement of the decoration units, altering their geometry can tune their spectral response. 
In our design framework, we assume that each decorator acts as an individual resonator, allowing for precise spectral control \cite{vilayphone_design_2024, babicheva_multipole_2021}. 
To investigate the spectral tunability, we conduct a simulation study comparing differently shaped dielectric resonators placed at identical positions (i.e., the metasurfaces are created from the same point pattern) on a lossless, dispersion-less dielectric substrate. 
As a reference, we also include a pattern generated by thresholding the same GRF image (see Methods for more details) to isolate the influence of the resonator shape. 
As shown in Fig. \ref{fig:fig3}a, we evaluate four configurations, from left to right: a thresholded GRF pattern (black) and a point pattern decorated with solid disks (r~=~100~nm, green), four-point star-shaped structures (r$_\mathrm{tip}$~=~64~nm, r$_\mathrm{indent}$~=~155~nm, purple)
and rings (r$_\mathrm{inner}$~=~75~nm, r$_\mathrm{inner}$~=~125~nm, orange).
All designs maintain a consistent filling fraction of $\sim 16\%$ to ensure a fair comparison by isolating the effects of the structural arrangement from those of material coverage.
The SDFs of all four 2D patterns are very similar (see Fig. \ref{fig:fig3}b), indicating that the form factor of the resonators has minimal effect on the spatial frequency distributions.
Notably, the SDFs of the four-point star and disk patterns are nearly identical, while that of the ring pattern shows a slight deviation due to the ring's form factor.
Given the similar SDFs and comparable filling fraction among the patterns, differences in the scattering spectra can be solely attributed to the nano-resonator. 

To study the angular and spectral dependence of scattering through the patterns, we calculate the far-field forward scattering for each pattern implemented as 200~nm-high resonators made of a dielectric of n~=~3.5 on a semi-infinite lossless substrate using finite difference time domain simulations.
Fig. \ref{fig:fig3}c shows the calculated scattered power fraction as a function of photon energy and in-plane momentum ($k_{//}$). 
The targeted angular scattering region given by the SDF is highlighted with dotted white lines. In all patterns, the scattered power is concentrated within this region.
However, the scattering intensity within the $k \in [10,20]~\upmu$m$^{-1}$ interval exhibits strong dependence on the photon energy, where the trend varies for the different resonator geometries. 
For the thresholded pattern, the scattered power fractions show a broadband feature centered at about 1.7 eV.
The scattering from the decorated metasurfaces displays two distinct scattering peaks, with energies that blue-shift as they move from the disks to the ring resonators. 
The spectral dependence is better quantified by integrating the scattered power over the  $k \in [10,20]~\upmu$m$^{-1}$ interval (Fig. \ref{fig:fig3}d).
By comparing the integrated power fractions with the forward scattering cross-sections of individual nano-resonators (Fig. \ref{fig:fig3}e), we show that spectral modulation originates from the excitation of resonant modes in the nano-resonator that favor forward scattering. 
We suspect slight spectra differences are likely due to lattice interactions \cite{zakomirnyi_collective_2019, babicheva_multipole_2021}.

\FloatBarrier


\FloatBarrier
\subsection{Superposition of metasurface designs in real and momentum space}
\begin{figure}
\begin{center}
\begin{tabular}{c}
\includegraphics[width=\textwidth]{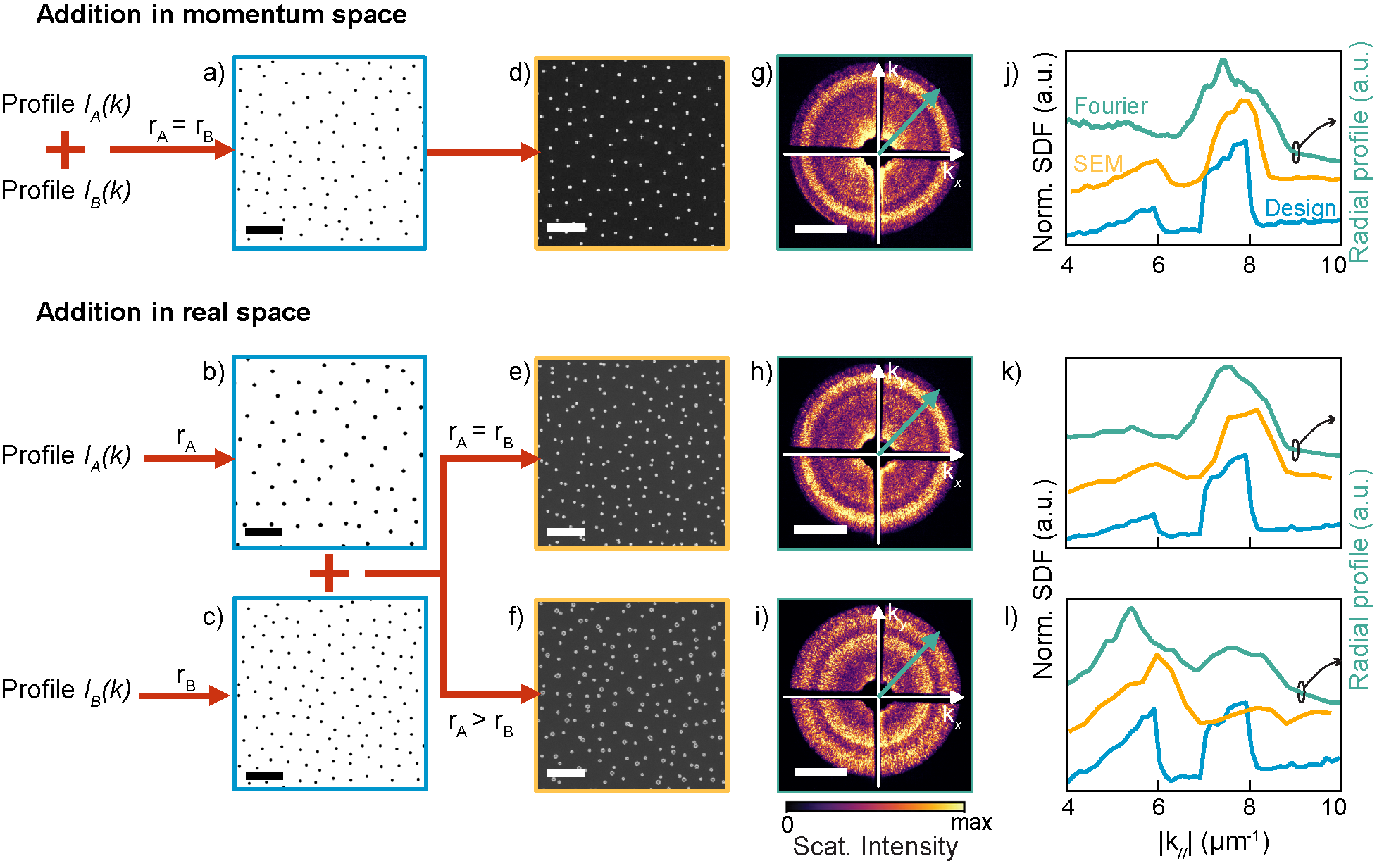}
\end{tabular}
\end{center}
\caption 
{\label{fig:fig5}
Generation of new metasurfaces based on the superposition of two patterns.
Profile A is defined as a rectangular function with a plateau between $k\in[4,6]$~$\upmu$m$^{-1}$; Profile B is defined as a rectangular function with a plateau $k\in[7,8]$~$\upmu$m$^{-1}$.
Columns from left to right: (a-c) Real space patterns created using our method for the different input profiles. The scale bar is 2~$\upmu$m.
(d-f) SEM images of fabricated Au pillars on Si substrate. The scale bar is 2~$\upmu$m.
(g-i) Fourier scattering microscopy images of the nanostructures obtained by illumination with unpolarized laser light filtered at $\lambda$~=~640~nm with 10~nm bandwidth. The colorbar indicates the scattering intensity where each image is normalized to its maximum.
(j-l) Extracted SDFs of the real space design and the SEM images, compared with the radially averaged profiles extracted from the  Fourier microscopy images in (g-i). The graphs are normalized to the maximum value in the depicted range and shifted vertically for clarity.
} 
\end{figure}

We now explore how more complex metasurfaces can be generated by superimposing two (or more) independently designed patterns. This strategy enables hybrid designs that combine distinct angular or spectral scattering features within a single device (layer).
Under the assumption of linear, non-interacting scattering contributions, metasurface designs can be combined either in the real space (by overlaying spatial patterns) or in the momentum space (by summing their target SDFs). We explore both approaches with the two input functions, $I_A(k)$ and $I_B(k)$, as previously introduced.

For adding patterns in the momentum space (Fig. \ref{fig:fig5}, top row), the combined SDF $I_{A,B}(k) = I_A(k) + I_B(k)$ is used to generate a point pattern subsequently decorated with identical disk-shaped resonators (r~=~80~nm, Fig. \ref{fig:fig5}a).
For adding patterns in the real space (Fig. \ref{fig:fig5}, middle, bottom row), two independent point patterns$-$derived separately from $I_A(k)$ and $I_B(k)$$-$are overlaid upon decoration with disks of either equal radii ( ($r_A$~=~$r_B$~=~80~nm) or different radii ($r_A$~=~113~nm, $r_B$~=~80~nm). 
The independent choice of a nanoresonator for each pattern opens an exciting avenue to target more complex scattering responses.
To experimentally validate this concept, we fabricate all metasurface variants as Au nanopillars on Si substrate (Fig. \ref{fig:fig5}d-f), resulting in radii of $\sim$195~nm and $\sim$260~nm, and a height of $\sim$300~nm.
The angular scattering properties are characterized using Fourier scattering microscopy centered at 640~nm wavelength (Fig. \ref{fig:fig5}g-i), resulting in double-ring scattering features for each metasurface, consistent with the composite input SDFs. 
The extracted radially-averaged profiles of the Fourier scattering (green), Fourier transform of the SEM images (orange), and the input designs (blue) are shown in Fig. \ref{fig:fig5}j-l). 
The strong correspondence among the three different curves for each metasurface confirms the regime of non-interacting structures. 

While both addition approaches produce the expected double-ring profiles, the relative scattering intensities differ. 
As noted in the previous section, features at larger wavevectors in the input profile lead to a higher density of points in these patterns, resulting in a systematic increase of intensity in the generated SDF as the wavevector increases \cite{yu_design_2017}. 
This is a direct consequence of our metasurface design method and explains the different intensities between the inner and outer ring radial profiles.
The effect becomes particularly evident when comparing the radially averaged profiles in Figs. \ref{fig:fig5}j,k). 
However, it can be counteracted by assigning larger disks to the lower-k lattice, which enhances scattering into low in-plane momentum regions.
As shown in Fig. \ref{fig:fig5}l), manipulation of the resonators' dimensions enables tuning the relative intensities of the inner and outer rings, which cannot be achieved through momentum-space addition.   
A detailed analysis of the nanoresonator geometry's effect on the relative scattering intensities is beyond the scope of this work and will be tackled in the future.
In summary, we observe that the relative intensities of the two peaks are impacted by the size of the decoration unit, offering an easy way to tune the scattering into different angular ranges.
Real-space addition combined with independent decoration enables the reproduction of superimposed angular features and tailoring their relative intensities, offering a new and powerful way of scattering control.

\FloatBarrier

\subsection{Simultaneous control over spectral and angular scattering response using real-space addition}

\begin{figure}[h!]
\includegraphics[width=\textwidth]{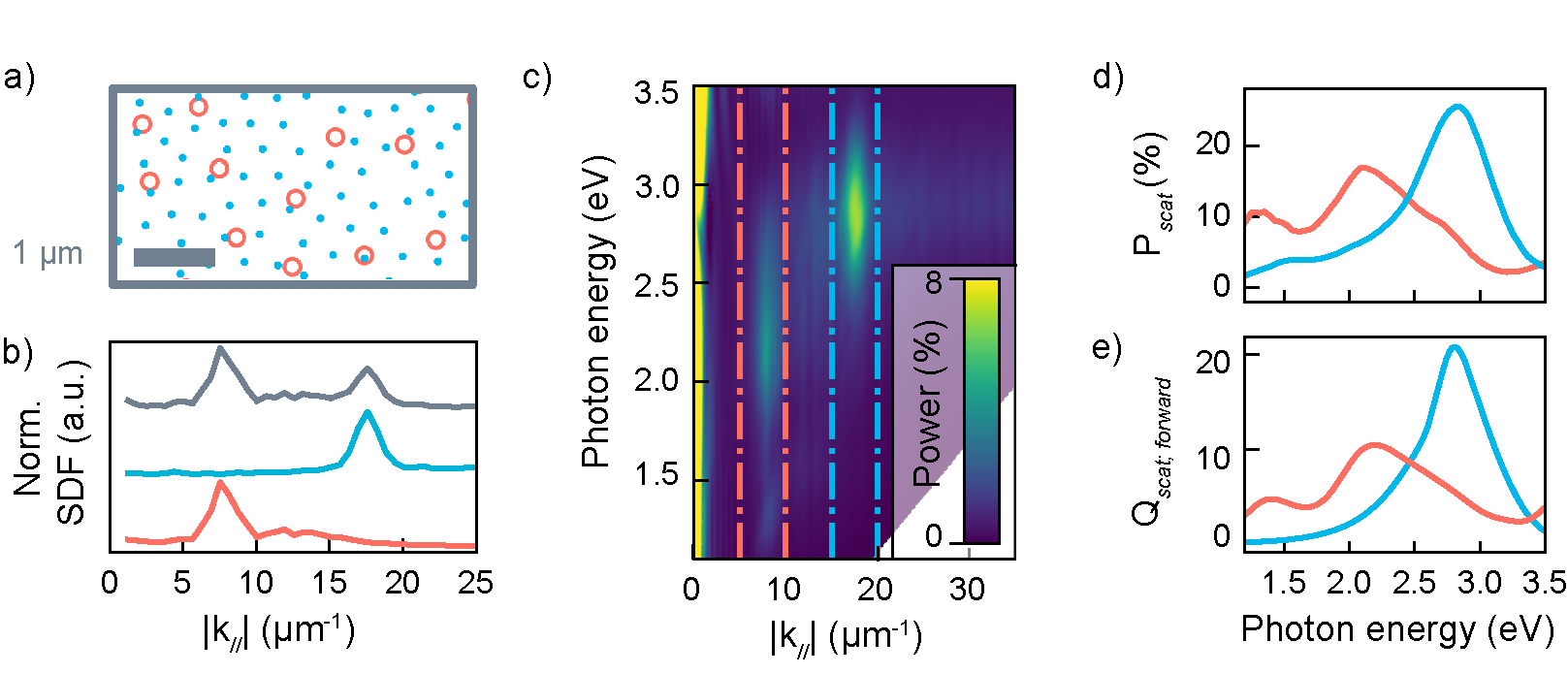}
\caption 
{ \label{fig:fig7} 
Far-field scattering for a pattern with different decoration units compared to the forward scattering for each building block.
(a) Summation of two patterns designed for different scattering ranges. 
The underlying subpatterns are highlighted in different colors: a point pattern designed for the range of $k \in [5,10]~\upmu$m$^{-1}$ is decorated with rings ($r_\mathrm{inner}$~=~75~nm, $r_\mathrm{outer}$~=~125~nm) and shown in red. Blue is a point pattern for the range of $k~\in~[15,20]~\upmu$m$^{-1}$ and decorated with a disk of r~=~50~nm. The patterns were added in real space into a single plane; the colors are for clarification only.
(b) Corresponding SDF from the different patterns. In red (blue), the SDF for the ring (disk) subpattern is given. Each subpattern was created with a triangular function of 5~$\upmu$m$^{-1}$ width and peak at the center. The SDF of the summation of patterns is shown in grey. 
Each profile is normalized to its maximum, and the patterns are shifted for clarity.
(c) Fraction of the incoming power scattered into the far-field for the pattern in (a) implemented as dielectric (n~=~3.5, h~=~200~nm) on a lossless, dispersion-less substrate of the same refractive index. The dash-dotted lines indicate the range for which each subpattern was designed in the respective color.
(e) Integrated power for each subpattern, i.e., the ring subpattern integrated for range $k~\in~[15,20]~\upmu$m$^{-1}$ (red) and the disk subpattern integrated for range $k~\in~[15,20]~\upmu$m$^{-1}$ (blue).
(f) Forward scattering cross-sections of the two different nano-resonators used to generate the pattern (red: ring-shaped, blue: disk-shaped).
For color: see online.
} 
\end{figure} 

Finally, we leverage the two distinctive features of our pattern generation approach to fully exploit its potential: on one hand, the possibility to decorate each position with a different scattering unit and, on the other hand, the flexibility to superimpose multiple point distributions in real space.
Designing a metasurface in this way enables simultaneous and independent control over the angular and spectral scattering responses.

As a proof of concept, we target two different angular regions centered at $k_1$~=~7.5~$\upmu$m$^{-1}$ and $k_2$~=~17.5~$\upmu$m$^{-1}$ in a computational analysis. For this, we generate two point distributions with our approach using a triangular input function with a width of 5~$\upmu$m$^{-1}$.
Each point distribution is decorated with a different shape: one with a ring and one with a disk (Fig. \ref{fig:fig7}a).
In Fig. \ref{fig:fig7}b), the corresponding SDFs are shown: blue for the sub-pattern with disk-shaped decoration units, which has features at higher angular frequencies; red for the sub-pattern with ring-shaped decoration units and features at lower angular frequencies; and gray for the real-space addition of the blue and red sub-patterns.
As expected, the summation of the two patterns retains the angular features of the two sub-patterns: the SDF (gray line) of the addition-pattern shows features at the same positions as the sub-patterns creating it (blue, red line).

Next, we simulate the angular and spectral scattering of a metasurface with this pattern consisting of a dielectric (n~=~3.5, h~=~200~nm) on a substrate of the same material with FDTD.
It is evident from Fig. \ref{fig:fig7}c) that the angular scattering is enhanced in the designed ranges (vertical dashed lines). 
In each interval, at higher or lower in-plane momentum ranges, the scattering intensities peak at different photon energies. 
Integrating the power scattered into the targeted k-space regions (Fig. \ref{fig:fig7}e) reveals enhanced scattering at spectral positions that correspond to the forward scattering of the single nano-resonator.
Here, the spectral scattering response is again clearly linked to the forward scattering of the resonator unit cell. 
This result highlights the versatility of our approach to tailoring the design of metasurfaces to specific applications by carefully adjusting the angular and spectral response.


\section{Conclusion}
\label{sect:conc}

In this work, we have introduced a non-iterative reverse design approach for correlated-disordered metasurfaces that allows simultaneously controlling the spectral and angular scattering of incident light.
It is based on a 1D representation of the Fourier space, the SDF, and uses blob detection to identify correlated-disordered points with given angular scattering properties. 
A subsequent re-decoration step allows for further tailoring of the optical response.
We fabricated the patterns as Au nanopillars on Si substrates and used Fourier microscopy to measure the angular scattering of simple pillar distributions.
The experimental results strongly support our claim that our binary pattern generation approach can be used to design metasurfaces with defined scattering properties.
In a complementary computational analysis based on FDTD calculations of dielectric resonators, we show that the forward scattering properties of the single decoration unit modulate the ensemble's spectral scattering response, yielding a design parameter that can be tailored exclusively to the spectral range of interest. 
One key advantage of our proposed approach is the possibility of superimposing two metasurfaces via real-space addition. 
This feature allows us to independently design angular and spectral responses for specific applications, as demonstrated experimentally and by simulations.
In this work, we introduce a straightforward method for designing metasurfaces that does not rely on iterations or optimizations and demonstrate how the angular and spectral characteristics of light scattering can be individually controlled.
While we acknowledge that implementing iterative optimization steps could further enhance the control over the angular scattering response, our method already provides flexible control over the point distribution.
To highlight the advantages of our method's re-decoration strategy (compared to established ones), we tune the spectral response by changing the resonator's geometry; however, the same concept can be extended to other properties, such as phase or polarization. 
All combined, we present a versatile toolbox for metasurface engineering.

\FloatBarrier

\newpage

\section*{Disclosures}
While preparing this work, the authors used ChatGPT to refine the code and Grammarly to edit the language. The authors reviewed and edited the content as needed and take full responsibility for the publication's content.

\section*{Code, Data, and Materials Availability}
The data and code are available upon request.

\section*{Acknowledgments}
This project has received funding from the European Union’s Horizon 2020 research and innovation programme under the Marie Skłodowska-Curie grant agreement no. 945363. 
The authors thank the EPFL Center for Imaging for helpful discussions, and EPFL CMi and CiMe for nanofabrication and characterization infrastructure. 
This work is part of the Dutch Research Council (NWO) and was partly performed at the research institutes AMOLF and ARCNL. The Advanced Research Center for Nanolithography ARCNL is a public-private partnership between the University of Amsterdam, Vrije Universiteit Amsterdam, University of Groningen, the Netherlands Organization for Scientific Research (NWO), and the semiconductor equipment manufacturer ASML.
This work used the Dutch national supercomputer Snellius with the support of the SURF Cooperative using grant no. EINF-8457 and grant no. EINF-9404.

\section*{References}

\bibliography{references}   
\bibliographystyle{spiejour}   

\newpage



\clearpage

\phantomsection
\renewcommand\thesection{S}

\section{Supporting Information}

\renewcommand{\figurename}{S}
\setcounter{figure}{0}

\FloatBarrier

\subsection{Implementation of point pattern generation using blob detection: Flow chart}
\begin{figure}
\begin{center}
\begin{tabular}{c}
\includegraphics[width=0.95\textwidth]{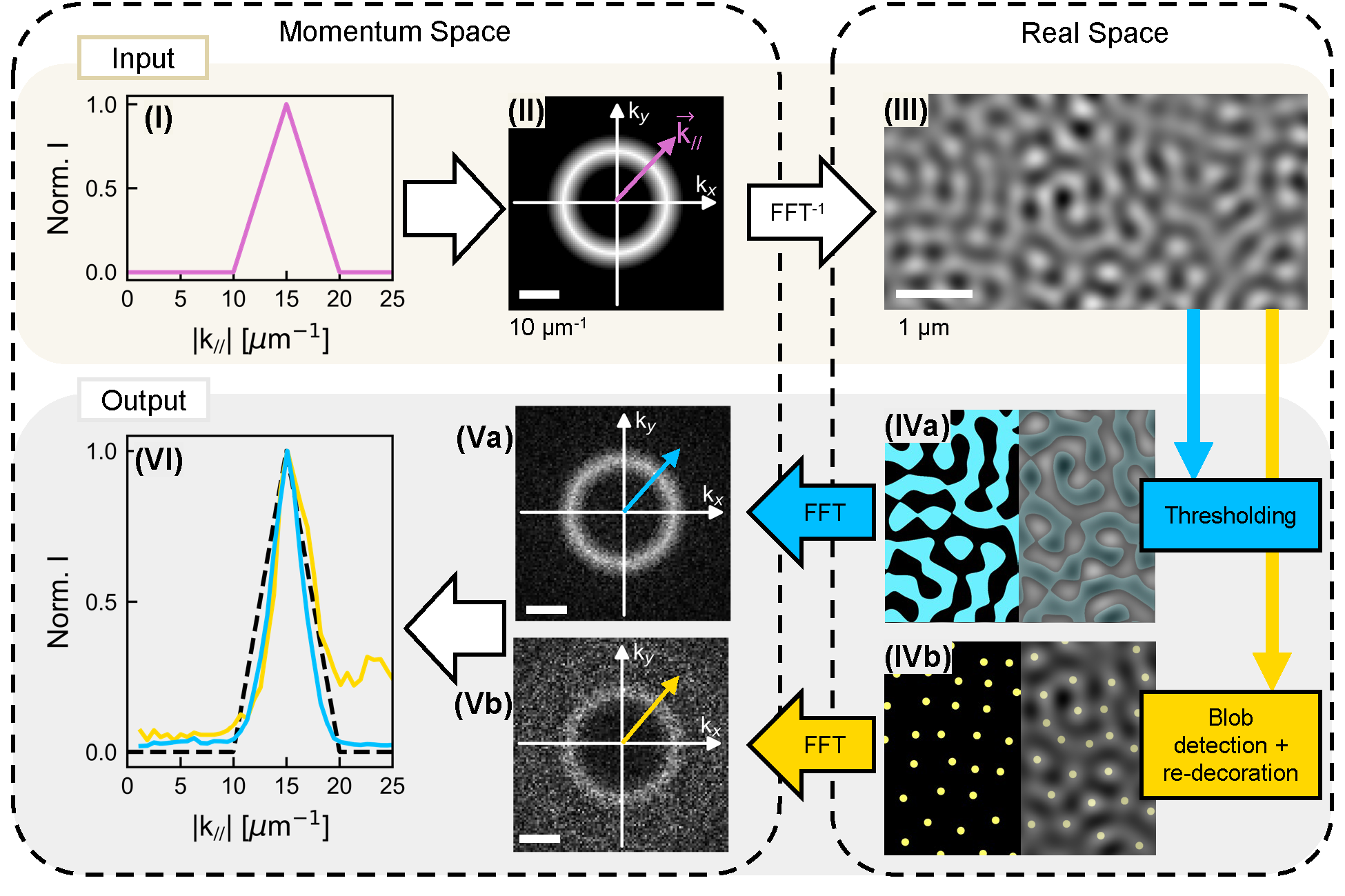}
\end{tabular}
\end{center}
\caption 
{ \label{fig:fig1}
Overview of the introduced point pattern generation method via blob detection and compared to thresholding.
Given an arbitrary input SDF (I) of the 1D radial profile, the 2D power spectral density (PSD) (II) is created in the momentum space. Each point is assigned a random phase.
(III) Inverse Fourier transform of the PSD, resulting in a correlated Gaussian random field (GRF). 
(IV) Two ways can be implemented to generate a two-phase pattern from this: 
(IVa) by defining a threshold (blue) or (IVb) by using a blob detection algorithm that outputs a set of coordinates that are then redecorated with a decoration unit (in this case, disks with $r\ =\ 50\ \mathrm{nm}$).
The PSD of the two-phase patterns is checked by FFT (Va,b) and their radially averaged profiles are compared to the input function (VI).
} 
\end{figure}

This pattern generation algorithm uses several steps, each with its own variables; the flowchart indicating the different steps can be found in \ref{fig:fig1}.
In the following part, we introduce the relevant variables for the generation in more detail and show the robustness of this approach to parameter variations.

Starting from the target SDF $I(\mathbf{k})$ (Fig. S\ref{fig:fig1}-I), the Fourier space is constructed arbitrarily for each point $\mathbf{k}$ with a particular amplitude $A_\mathbf{k}$ and phase $\phi_\mathbf{k}$.
The size of the momentum space is defined by the target real space size, represented by the number of pixels and pixel size of the image. This can be chosen arbitrarily and is only limited by the available computational memory.
The amplitude $A_\mathbf{k}$ is chosen from a random value between $|I(\mathbf{k}) \pm g \cdot I(\mathbf{k})|$. 
The range of values (given by $2 g$) is arbitrary, but care must be taken to preserve a difference.
The phase $\phi_\mathbf{k}$ is chosen randomly between $\pm\pi$.
To preserve the system's physicality, only one half of the momentum space is constructed, and the other half is defined as the complex conjugate of the point-symmetric counterpart.
In the case of the input function as in (Fig. S\ref{fig:fig1}.-I), the result (Fig. S\ref{fig:fig1}.-II) is a characteristic ring in momentum space with a triangular cross-section.
An inverse Fourier transform yields a correlated Gaussian random field (GRF) (Fig. S\ref{fig:fig1}-III) in which the pixel values follow a Gaussian distribution.
In-plane spatial correlations are evident in high- or low-value clusters in the real space.

To achieve a binary (or ternary) pattern, classic GRF modeling consists of level-cutting (thresholding) the GRF by assigning each pixel above a particular value $t$ to one level and each pixel below $t$ to another level, resulting in a two-level pattern with channel-type morphology (Fig. S\ref{fig:fig1}-IVa).
In this work, we have introduced a blob detection step to identify points of higher correlation in the GRF, which are subsequently re-decorated with specific shapes (Fig. S\ref{fig:fig1}-IVb). 
There are different ways to identify the position of the blobs. For our approach, we calculated the Laplacian of Gaussians of the image because it provides the highest read-out accuracy \cite{van_der_walt_scikit-image_2014}.
The function calculates the Laplacian of the image blurred by a Gaussian with a determined standard deviation $\sigma$ (in px) for a certain number of $\sigma$. It also applies a lower threshold for detecting blobs. 
As the size of the detected blobs is related to the minimal (maximal) $\sigma$ values provided, we determine the expected sizes of the blobs by looking at the highest (lowest) expected spatial frequencies from the input SDF and convert them to the corresponding pixel size.
The lower threshold for blob detection is set to 1\% of the maximum of the image values, and no overlap of blobs is allowed.
Comparing the power spectra of the real space structures (Fig. S\ref{fig:fig1}-Va, b) and SDFs with the input (Fig. \ref{fig:fig1}-VI) shows that most information is retained for both approaches.

\FloatBarrier
\subsection{Comparison with GRF modelling and versatility}

\begin{figure}
\begin{center}
\begin{tabular}{c}
\includegraphics[width=0.95\textwidth]{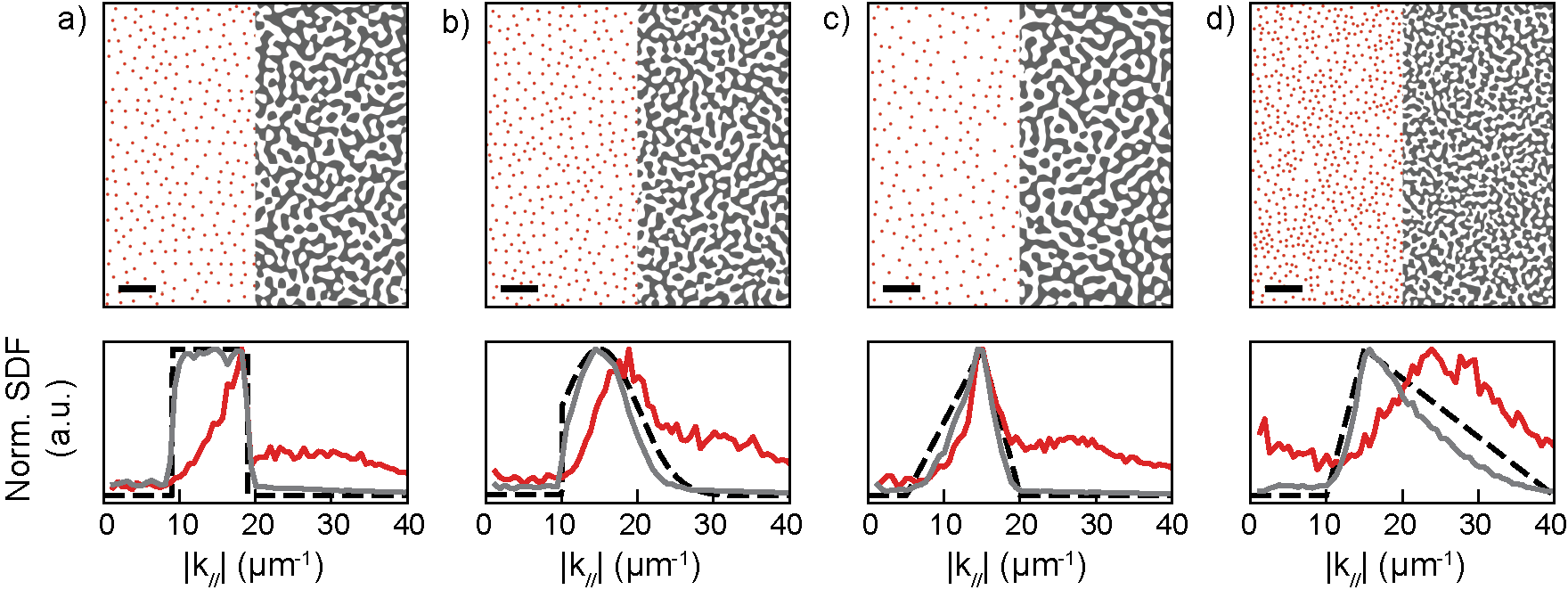}
\end{tabular}
\end{center}
\caption 
{Comparison of resulting SDFs for thresholded and point patterns generated with this method for the same input SDF: (a) rectangular, (b) truncated Gaussian, (c, d) triangular asymmetric.
For each input, the patterns have a size of 5000x5000~px and 2~nm/px resolution. 
The point patterns (red, left) have a decoration unit of r=50~nm, and the thresholded patterns (gray, right) a filling fraction of 50\%. 
The scale bar corresponds to 500~nm.
The lower graphs compare the input (black dashed line) with the extracted SDFs in the respective color.
\label{fig:si2}
} 
\end{figure} 

To show the versatility of this approach and how it compares to standard GRF modelling, we take different types of input SDFs, generate binary point patterns and binary thresholded patterns with standard parameters, and compare them to the resulting SDFs of the patterns. 
We compare four different types of input functions and depict the results in Figure \ref{fig:si2}.
Each subfigure shows the generated point patterns with disk-shaped decoration units of r~=~50~nm (left, red), and of the thresholded pattern with a filling fraction of 50\% (right, grey) for the same GRF. 
At the bottom of each subfigure, the input profiles (black dashed line) and the resulting SDFs (red, grey solid lines respectively) are shown.

Overall, the thresholded patterns retain the spatial information the input SDF gives better, independent of its shape. This is reasonable because the thresholded patterns with spinodal-like features carry more spatial details lost in the point read-out.
For the re-decorated point patterns, differences at higher spatial frequencies are also influenced by the choice of decoration unit.

\FloatBarrier
\subsection{Scalability and impact of decoration unit}

\begin{figure}
\begin{center}
\begin{tabular}{c}
\includegraphics[width=0.95\textwidth]{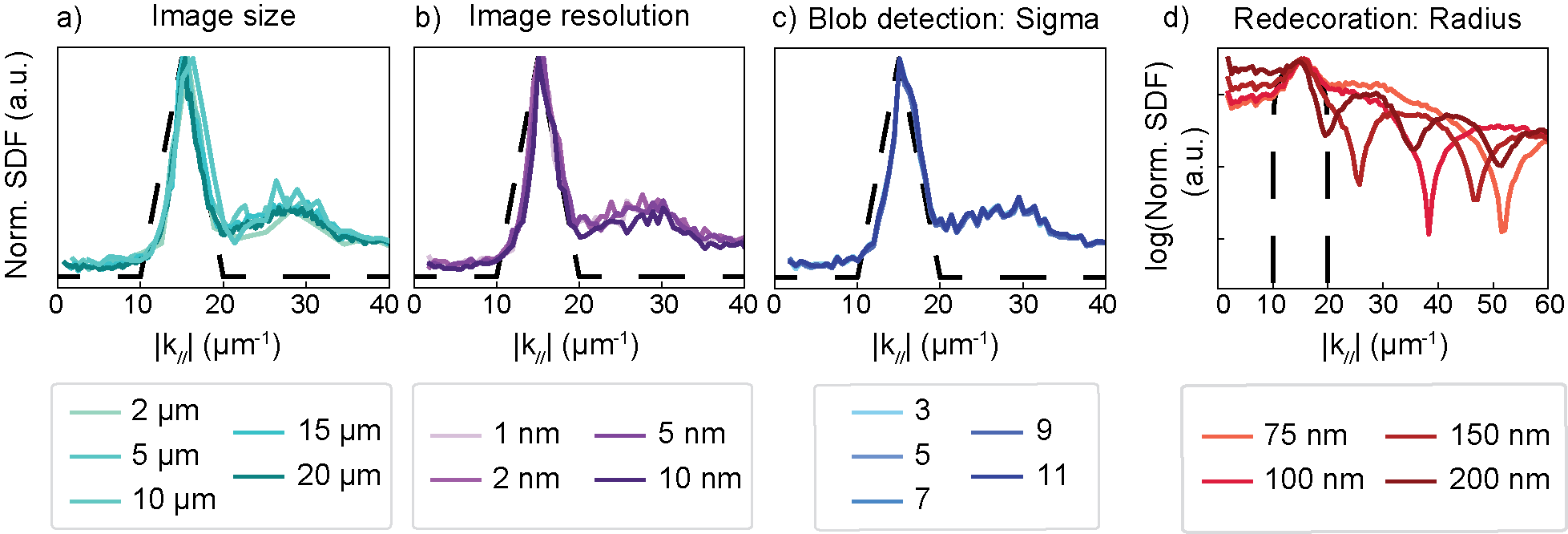}
\end{tabular}
\end{center}
\caption 
{
Effect of change in variables on generated SDFs for the same input and standard parameters. 
Standard parameters are 10~$\upmu$m image size, 2~nm/px image resolution, three (3) sigma for the blob detection, and r~=~50~nm as decoration unit.
The solid black dashed line shows the input function (normalized).
(a) Variation of the image size, i.e., the number of pixels in the square image.
(b) Variation of the resolution, i.e., nm/px.
(c) Variation of the number of standard deviations used in the blob detection algorithm. 
(d) Variation of the chosen radius for a disk-shaped decoration unit.
\label{fig:si1a}
} 
\end{figure}

To understand the impact of different variables on the accuracy of the point patterns generated with our approach, we calculate point patterns using a triangular input function for various parameters and compare the resulting SDFs with each other. 
Specifically, we look at the image size (i.e., the number of pixels for an image), image resolution (i.e., nm/px) on the input side, the number of standard deviations for the blob detection on the read-out side, and lastly, the impact of the choice of radius on the redecoration side. 

Figure S\ref{fig:si1a} shows how the SDF for the generated point patterns varies when one variable is changed from the standard parameters. 
From figure S\ref{fig:si1a}, a, b, it can be seen that image size or resolution only slightly affects the SDF of the generated image of the pattern, and the spatial features from the input are well represented. 
As the image size and resolution determine the number of pixels in the real and momentum space, they directly affect the memory consumption of the pattern generation.
It is recommended to choose a resolution as big as possible to save memory, but in general, point patterns of arbitrary lateral dimensions can be generated.

The blob detection readout uses the "Laplacian of Gaussians" to identify the blobs. Each Gaussian is calculated with a determined standard deviation, sigma. The user can define the number of standard deviations, i.e., the number of images, used for the readout algorithm. 
It might be tempting to assume that increasing the number of images can increase the readout accuracy (which, at the same time, increases the algorithm's time). 
We run the blob detection algorithm on the identical GRF image with an increasing number of sigmas.
From the resulting, nearly overlapping SDFs in Fig. S\ref{fig:si1a}c, it can be deduced that the blobs detected are roughly identical.
Therefore, it is not necessary to consider more standard deviations to extract the information relevant to re-creating the spatial features in a point pattern, saving computation time and memory.

At spatial frequencies corresponding to the diffraction pattern of a single decoration unit, the SDF is impacted by it. 
This directly follows from our choice to use the SDF (derived from the power spectral density of the output pattern) as representation of the real space patterns, which is effectively the convolution of the lattice geometry and the decoration unit, i.e., the product of the lattice's structure factor and the decoration unit's form factor.
In Fig. S\ref{fig:si1a}d, it is obvious how the disk-shaped decoration unit's radius impacts the binary pattern's SDF. 
For increasing radius, the bounce-like feature originating from the disk's form factor shifts to lower wavevectors, in agreement with the general relationship between real and Fourier space: bigger structures in real space have a narrow extension in Fourier space, and vice versa.

It should be noted that the desired outcome defines the momentum space that can sensibly be designed with this point-pattern generation method.
The density of points determined by the blob detection increases when moving the defining features of $I(\mathbf{k})$ towards larger spatial frequencies, as the points of higher correlation become closer to one another. 
This limits the maximum decorator size that can be placed at each point without overlapping with neighboring ones.
Therefore, from an application point of view, it is necessary to consider the trade-off of physical considerations like the targeted momentum space and the desired shape and dimensions of the decoration unit.

\end{spacing}
\end{document}